\documentclass[letterpaper,10pt]{article}
\usepackage{osameet2}

\usepackage{amsmath,amssymb}

\usepackage[dvips,colorlinks=true,bookmarks=false,citecolor=blue,urlcolor=blue]{hyperref} 

\begin{document}

\title{Observation of Spatiotemporal Chaos Induced by a Cavity Soliton in a Fiber Ring Resonator}

\author{Miles Anderson, Fran\c{c}ois Leo, St\'ephane Coen, Miro Erkintalo, and Stuart G. Murdoch}
\address{\hspace*{-10mm} Dodd-Walls Centre and Department of Physics, The University of Auckland, Private Bag 92019, Auckland 1142, New Zealand\hspace*{-10mm}}
\email{m.erkintalo@auckland.ac.nz}

\begin{abstract}
We report on the experimental observation of temporal cavity soliton destabilization via spatiotemporal chaos in a coherently-driven optical fiber ring resonator. Numerical simulations and theoretical analyses are in good agreement with experimental observations.
\end{abstract}

Coherently-driven passive nonlinear optical cavities exhibit rich dynamical behaviours, including bistability and pattern formation~\cite{lugiato_book}. Although historically such dynamics have predominantly been investigated in the context of spatially \emph{diffractive} cavities~\cite{ackemann_fundamentals_2009}, recent years have witnessed the research focus continuously shifting towards \emph{time-domain dispersive} systems such as fiber ring resonators~\cite{leo_temporal_2010,leo_dynamics_2013, jang_temporal_2015} and monolithic microresonators~\cite{herr_temporal_2014, chembo_spatiotemporal_2013}. Besides fundamental interest, the shift has been driven by applications: driven fiber-ring resonators enable high-performance optical buffers~\cite{leo_temporal_2010, jang_temporal_2015}, whilst microresonators permit creation of ``Kerr'' frequency combs~\cite{herr_temporal_2014, kippenberg_microresonator-based_2011}. Central to both applications are localized nonlinear structures first observed in 2010 in a fiber cavity~\cite{leo_temporal_2010}. Known as temporal cavity solitons (CSs), they are pulses of light that can circulate indefinitely in the resonator, and they correspond to steady-state solutions of the underlying AC-driven nonlinear Schr\"odinger equation (NLSE)~\cite{leo_temporal_2010}.

\looseness=-1 Temporal CSs exist for particular pump-resonator parameters only, and under conditions of strong cavity driving, they can exhibit oscillatory ``breathing'' behaviour as well as other complex instabilities~\cite{leo_dynamics_2013}. In particular, it has been theoretically predicted~\cite{leo_dynamics_2013} that a CS can act as a ``seed'' for a type of spatiotemporal chaos of the AC-driven NLSE that was first identified in numerical experiments on plasma physics in 1985~\cite{noziaki_chaotic_1985}. Enabled by its emerging under conditions of continuous wave (cw) bistability, such spatiotemporal chaos can exhibit temporal localization~\cite{leo_dynamics_2013}, differentiating the phenomenon from the fully de-localized chaotic modulation instability (MI) that has recently attracted attention in the context of microresonator frequency combs~\cite{erkintalo_coherence_2014}. However, despite being rooted in a long history of theoretical analyses~\cite{noziaki_chaotic_1985, leo_dynamics_2013}, the pertinent dynamics has so far eluded experimental observation. Here, we report on the experimental observation of spatiotemporal chaos induced by temporal CSs in a coherently-driven passive fiber ring resonator.

\looseness=-1 The fiber ring resonator used in our experiments is similar to that in~\cite{jang_temporal_2015}. However, that study relied on a cw laser to drive the resonator, which is not suitable for our needs. Indeed, for this resonator, spatiotemporal chaos is predicted only when driven with several watts of power, which is difficult to realise in pure cw. To alleviate this issue, we follow an approach similar to that in~\cite{copie_competing_2015}, i.e., we drive the resonator with flat-top nanosecond pulses, obtained by intensity modulating a narrow linewidth, tunable cw laser. The intensity modulator is synchronised to the cavity roundtrip time, ensuring that the pulsed nature of the driving field does not materially alter the dynamics. Before injection into the cavity, the nanosecond pulses are amplified such that their quasi-cw peak power is about $P_\mathrm{in} = 7~\mathrm{W}$, yielding a normalized driving strength $S = 5.6$. (The normalization is the same as in~\cite{leo_temporal_2010, leo_dynamics_2013}, i.e., $S=(P_\mathrm{in}\gamma L\theta/\alpha^3)^{1/2}$, with $\gamma$ the nonlinearity coefficient, $\theta$ the input coupler power transmission coefficient and $\alpha$ half the percentage power lost per roundtrip.)

\begin{figure}[t]
    \includegraphics[width=\textwidth]{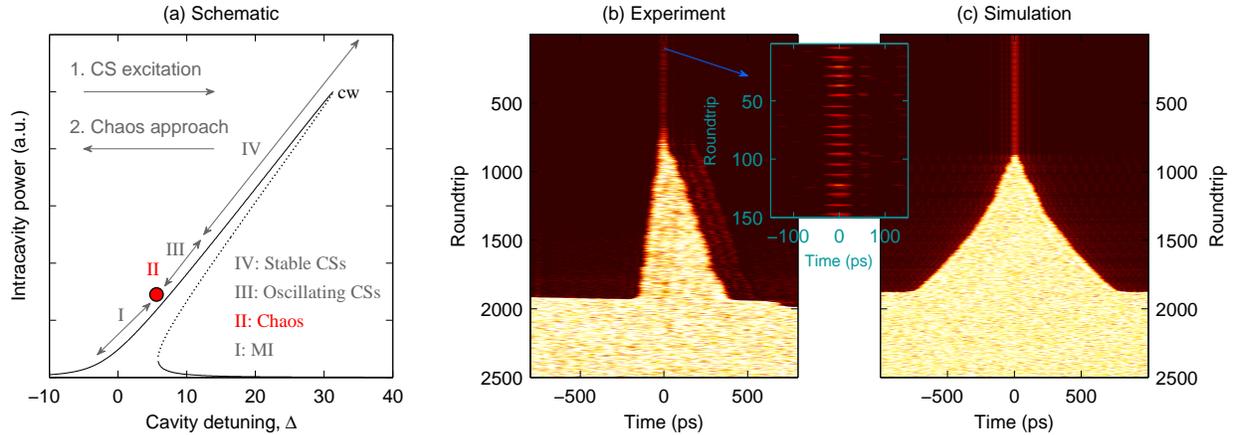}
    \caption{\looseness=-1(a) Schematic of dynamical regimes. Black curve is the bistable cw cavity response (dotted curve represents homogeneously unstable cw solutions). We first excite \emph{stable} CSs by increasing $\Delta$. We then reduce $\Delta$ to a region of interest, and record the cavity output in real time. (b) Experimental results, showing the cavity output over 2500 roundtrips when the detuning is reduced (from top to bottom) around the boundary between oscillating CSs and spatiotemporal chaos ($\Delta\approx 6$). (c) Corresponding results from numerical simulations. Bright (dark) colors indicate high (low) intensity.}
    \label{Fig1}
    \vskip -10pt
\end{figure}

For constant driving strength, the dynamics of our system depend only on the detuning of the driving field (with frequency $\omega_\mathrm{p}$) from the closest cold cavity resonance at $\omega_0$~\cite{leo_temporal_2010, leo_dynamics_2013}. Referring to a normalized detuning $\Delta \approx \mathcal{F}t_\mathrm{R}(\omega_0-\omega_\mathrm{p})/\pi$, with $\mathcal{F}$ the cavity finesse and $t_\mathrm{R}$ the roundtrip time, Fig.~\ref{Fig1}(a) schematically illustrates the different theoretically predicted dynamical regimes for our experimental $S = 5.6$, superimposed with the cw cavity response (black curve). Spatiotemporal chaos is to be expected over a narrow range of low detunings, whilst oscillating and non-oscillating (i.e., stable) CSs manifest themselves over wider ranges and for larger detunings.

In the beginning of our experiment, we excite temporal CSs by scanning the initially blue-detuned driving laser across a cavity resonance (i.e., we slowly increase $\Delta$). This results in the spontaneous creation of stable CSs~\cite{herr_temporal_2014, luo_spontaneous_2015}. Once the CSs have been excited, we reverse the direction of the laser scan, i.e., we begin to reduce the detuning $\Delta$. We then considerably slow down the scan speed around a detuning of interest, and record a long real time signal at the cavity output on a fast oscilloscope. This general approach enables a detailed examination of how the CS behaviour changes with the cavity detuning. Here we focus on spatiotemporal chaos [see Fig.~\ref{Fig1}(a)], yet note that the technique also allows us to investigate other complex oscillatory instabilities as well as excitability.

Example measurements for detunings in the spatiotemporal chaos regime are shown in Figure~\ref{Fig1}(b). Here we display a concatenated sequence of oscilloscope traces, recorded over 2500 roundtrips, as the detuning is slowly reduced (top to bottom) around $\Delta \approx 6$. Up to the 700th roundtrip, we observe a single localized CS that exhibits high-contrast ``breathing'' oscillations (see inset)~\cite{leo_dynamics_2013}. Then suddenly, the CS destabilizes entirely, transforming into a fluctuating field that is bounded in time, yet expands from roundtrip-to-roundtrip. Given that the cavity photon lifetime is only about 3 roundtrips, the front expansion can be appreciated to occur over slow dynamical timescales. Moreover, the co-existence of the chaotic state with the homogeneous cw state that exists beyond its boundaries agrees very well with characteristics predicted for spatiotemporal chaos~\cite{leo_dynamics_2013}. When the detuning is reduced beyond the point where the stable lower cw state ceases to exist [around $\Delta\approx 6$, see Fig.~\ref{Fig1}(a)], the entire field transforms into a non-localized unstable MI state, allowing for direct distinguishing of the two chaotic regimes. These experimental observations are in excellent agreement with numerical simulations of the AC-driven NLSE, as can be appreciated from Fig.~\ref{Fig1}(c).

In conclusion, we have reported on an experimental technique that allows for various temporal CS instabilities to be explored. This technique has enabled us to directly observe, for the first time to our knowledge, spatiotemporal chaos of the AC-driven NLSE. Besides fundamental interest, we expect our results to be of relevance to microresonator frequency comb experiments that typically operate in the strong driving regime.

\end{document}